# Generalizing Co-operative Evaporation in Two-Dimensional Droplet Arrays

Khushboo Pandey[1], Sandeep Hatte[2], Keshav Pandey[3], Suman Chakraborty[4], and Saptarshi Basu[1, 3*]

[1] Interdisciplinary Center for Energy Research (ICER), Indian Institute of Science, Bangalore 560012, India

[2]Department of Mechanical Engineering, Virginia Polytechnic Institute and State University, Blacksburg, 24061, USA

[3] Department of Mechanical Engineering, Indian Institute of Science, Bangalore 560012, India

[4]Department of Mechanical Engineering, Indian Institute of Technology Kharagpur, Kharagpur 721302, India

*Corresponding author: sbasu@iisc.ac.in

## Abstract

*Sessile droplets exposed to an incipient condition lead to an inevitable loss of mass, which is critical in many practical applications. By considering an arbitrarily configured two-dimensional array of droplets, here we provide a simple generalized theoretical limit to their lifetime in an evaporating state. Notwithstanding the geometrical and physical complexity of the effective confinement generated due to their cooperative interactions, we show that the consequent evaporation characteristics may be remarkably insensitive to the topographical details of the overall droplet organization, for a wide range of droplet-substrate combinations. With subsequent deployment of particle-laden droplets, however, our results lead to the discovery of a unique pathway towards tailoring the internal flows within the collective system by harnessing an exclusive topologically-driven symmetry breaking phenomenon, yielding a new strategy of patterning particulate matters around the droplet array.*

Evaporation of sessile droplets **(Stauber et. al. 2014; Sáenz et. al. 2015; Schofield et. al. 2018; Ghasemi & Ward, 2010; Xu & Choi, 2012; Chen et al., 2012)** is ubiquitous to our day-to-day experiences, including the formation of coffee-rings, ink-stains, as well as many commercial processes including spray painting, ink-jet printing, crop spraying, coating of seeds or tablets, spray cooling and spray drying. In most of these practical scenarios, the droplet arrangement can be best represented by topologically varying two-dimensional structures (2D arrays), resulting in obvious complexities in describing their dynamical characteristics by appealing to a universal physical paradigm.

The proximity of neighboring droplets in a multi-dimensional droplet array creates localized regions of vapor accumulation by saturating the interstitial voids. Consequently, an 'effective' vapor mediated confinement is established around the individual droplets. The distribution and extent of vapor accumulation leads to further modifications in the flow dynamics in and around the droplet, culminating in alterations in the global rate of evaporation, thereby influencing the evaporation dynamics of the droplet array in an intriguing manner **(Dugas et. al. 2005)**. Laghezza et. al. **(2016)**, through experiments and numerical simulations, showed that a system of interacting droplets results in increase in the dissolution lifetime of the central droplet. However, their formulation could not predict the droplet dissolution lifetime increment as a function of the droplet array configuration. Toledano et al. **(2005)** presented an analytical model for the cooperative evaporation of volatile droplet arrays. This approach, however, does not quantify the droplet evaporation lifetime. Carrier et al. **(2016)** formulated the droplet evaporation timescales by considering a group of interacting droplets as one flat superdrop. However, their asymptotic model loses its predictive capabilities under the limiting conditions of sufficiently sparse (lower vapor mediated interactions) distribution of sessile droplets.

An extensive review of literature reveals that despite significant advancements in the understanding of evaporation of single droplets or linear droplet arrays **(Laghezza et al. 2016; Tolédano, et. al. 2005; Carrier et al., 2016; Pradhan & Panigrahi, 2015; Hatte et. al. 2019; Bansal et. al. 2017a; Bansal et. al. 2017b; Shaikeea & Basu, 2016)**, generalization of the physics of cooperative evaporation of droplet arrays of arbitrary topology remains elusive. This deficit stems from the complexities in the possible depiction of a universal global scenario of the droplet evaporation characteristics, amidst the locally altered internal flows and evaporation patterns as a consequence of topographical modifications in the droplet array, wettability variations of the interfacing substrate, and participation of included particulate matters (like nanocolloids). Summarily, the following open questions remain to be addressed: (a) despite obvious symmetry breaking perspectives in disordered arrays of droplets and other diversities in droplet-substrate combination and droplet constitution, does any universal theoretical depiction exist for an arbitrary array of droplets under collective evaporation, (b) If so, how is the same related to the flow dynamics inside the droplets?

Here, we depict a generalization in the evaporation lifetime of a two-dimensional droplet array, irrespective of variations in the intrinsic wettability of the substrate, inclusion of particulate matters, and localized alteration in the droplet configuration topology. Our simple theoretical depiction of the extent of the vapor confinement borrows analogies from Voronoi tessellation **(Aurenhammer, 1991; Raju et. al. 2018; Du et. al. 1999; Fedorets et. al. 2017a; Kumar & Kumaran, 2005; Fedorets et. al. 2017b)**; a concept not traditionally employed for addressing droplet evaporation dynamics. Voronoi diagrams provide a definition of geometric neighbors. The seeds (droplets in the current scenario) sharing a common Voronoi edge are geometric neighbors **(Kumar & Kumaran, 2005)** that are responsible for the vapor confinement. Our results reveal that the droplet lifetime is remarkably independent of the relative positions of the surrounding droplets and the asymmetric nature of the droplet array and is uniquely described by the Voronoi tessellation area (Fig. 1a). Interestingly, even though the global evaporation lifetime is insensitive to the asymmetric droplet spacing in an array, the local evaporation flux does get altered significantly. This spatial alteration of evaporation flux brings out unique directional characteristics in the internal flow structure due to symmetry breaking phenomenon, leading to preferential deposition of particles around the droplet, in case the system is particle-laden. This leads to the discovery of a novel pathway of tailoring particle deposition patterns as mediated by externally controllable features of the internal flow structures within the droplet array.

For illustration, without sacrificing generality, $5 \times 5$ two-dimensional droplet array of DI water ($2.0 \pm 0.1 \mu l$) is printed over the substrates using the Precore Flowline SP FL500 syringe pump. The evaporation lifetime of the center droplet (surrounded by six adjacent droplets) is observed under an optical microscope (at $5X$ zoom, **Figure S1 supplementary material**). Experiments are carried out at $25 \pm 2\,°C$ and relative humidity of $45 \pm 2\,\%$. The experiments are conducted for multiple relative spacing between the individual droplets. Substrates of multiple intrinsic contact angles are utilized; Polydimethylsiloxane (PDMS, contact angle $\boldsymbol{\theta \sim 110° \pm 2°}$ ), Glass ($\boldsymbol{\theta \sim 35° \pm 2°}$ ), and Gas Diffusion Layer (GDL, $\boldsymbol{\theta \sim 135° \pm 2°}$ ).

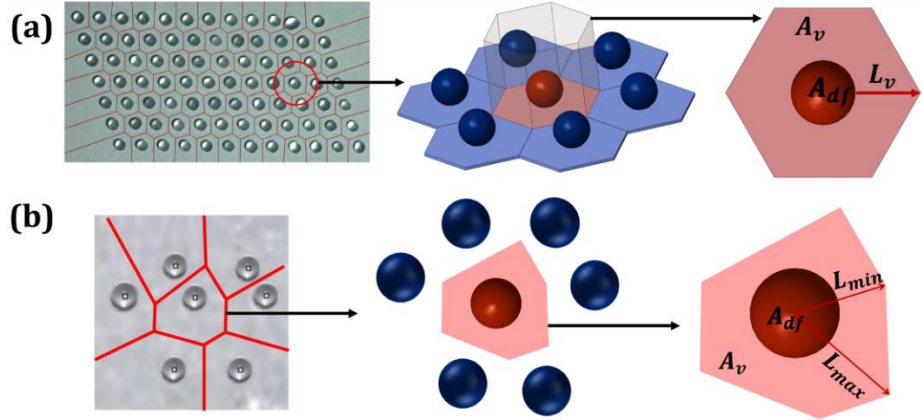

**Figure 1.** Voronoi tessellation of the (a) symmetric, and (b) asymmetric droplet array arrangement. Area of the Voronoi cell, $A_v$ quantifies the of vapor accumulation for a confined droplet of frontal area $A_{df}$ (**Supplementary video**).

During the initial transient phase, the evaporation front of each droplet moves radially outward without the recognition of the neighboring droplets (**Fig. 1**). At a certain spatio-temporal scale, these evaporation fronts merge, establishing steady state vapor confinement. In an effort to bring out the role of neighboring droplets on the same, we bring in perspectives from the concept of Voronoi tessellation **(Aurenhammer, 1991) (for details see Supplementary file)**. The Voronoi tessellation decomposes the plane into unit cells which are generated from the seeds/generator **(Raju et. al. 2018)**. In the present context, the confined droplets are the Voronoi generators and the respective Voronoi cells are the spatial interpretation of the vapour field of individual droplets (**see supplementary video**). In other words, a Voronoi diagram encodes the proximity information which in the present case helps to quantify the extent of vapour confinement (**Figure 1**). For symmetric 2D arrays, the unit cell forms a regular hexagon. Due to the symmetrical placement of the droplets, the Voronoi partitioning corresponds to the Centroid Voronoi Tessellation (CVT **(Du et. al. 1999)**) with area $A_v$ (**Figure 1 (a)**). However, asymmetric placement of droplets generates irregular hexagons (**Figure 1 (b)**). Note that here we neglect the transient timescales required to establish the steady state vapour confinement. The timescale necessary for the vapour to traverse from the droplet surface to the vapour confinement ($L_v$) lengthscale can be given as, $t_{trans} \sim L_v^2/D$, $D$ is the mass diffusivity of the water vapor in air. As a consequence, $t_{trans} \sim O(10^{-1} - 10^{-2})$ s $\ll t_c$ ; lifetime of confined droplet $\sim O(10^3)$ s.

The evaporation history of confined droplets (Figure S2, supplementary material) indicates three things; First, the confined droplets exhibit slower evaporation rate than an isolated droplet. Secondly, the respective evaporation rates are inextricably and uniquely linked with the extent of confinement, dictated by an exclusive confinement ratio $A_{df}/A_v$, where $A_{df}$ is the droplet frontal area from the top-view and $A_v$ is the area of the Voronoi cell. Thirdly, droplet lifetimes are independent of the degree of asymmetry, for the same $A_{df}/A_v$. However, the internal flow fields manifest modifications with both $A_{df}/A_v$ and the degree of asymmetry (Fig. 2). The asymmetry parameter is defined as the ratio of maximum to minimum length from hexagon center to the vertex, $L_{max}/L_{min}$ **(Fig.1(b))**.

**Fig.2 (a)** shows the bottom view flow field of an isolated sessile droplet. For a sessile droplet on a hydrophobic substrate, the maximum and minimum evaporation fluxes occurring at the apex region and three phase contact line result in a buoyancy driven double toroidal flow pattern **(Dash et. al. 2014)**. This results in radially outward flow, as seen from the bottom view (**Fig. 2(a)**), with an average velocity $\sim 8 \, \mu m/s$. In the present study, vapor mediated interaction reduces the average evaporation flux values and consequently, the average velocity of the internal flow field (**Fig.2 (b) and (c)**). For a minimum confinement ($A_{df}/A_v = 0.04$), the average velocity is observed to be $\sim 28\%$ lower while for the maximum confinement ($A_{df}/A_v = 0.21$) it is $\sim 42\%$ lower than that of an isolated sessile droplet case. **Figure 2(b)** follows that for a symmetric configuration, $L_{max}/L_{min} = 1$, the internal flow pattern remains radially outward as observed from the bottom view for both the extreme confinement values ($A_{df}/A_v = 0.04$ and $A_{df}/A_v = 0.21$). However, for $A_{df}/A_v = 0.21$ with $L_{max}/L_{min} = 2$, an asymmetric distribution of evaporation flux alters the internal flow pattern resulting to a unidirectional flow (**Figure 2(c)**). It indicates that the flow emerges from the region of higher extent of vapor accumulation (**Region I**) to the region of lower extent of vapor accumulation (**Region II**). Asymmetric droplet patterns, remarkably, do not affect the flow pattern for the minimum confinement ratio.

The local flow asymmetries are an artifact of the relative positions of the droplets for a given global confinement ($A_v$). As an example in **Fig. 2c**, Region II experiences lower vapor concentration or lower confinement compared to Region I. Consequently, the evaporation flux is higher from the side of the droplet facing Region II compared to Region I. The polar variation in evaporation flux, therefore, leads to directional flow **(Fig. 2c)**. The flow field becomes increasingly directional with increase in the asymmetry parameter, while the global evaporation timescale remains largely unaltered for the same $A_{df}/A_v$. The cumulative evaporation flux (Region I +Region II) of an asymmetric arrangement thus remains similar to that of the symmetric case.

Utilization of flow asymmetries, however, can be a novel approach towards harnessing preferential migration of particulate matter around particle-laden droplets. This proposition has been verified by a simple experiment of drying coffee droplets. The thickness and height of the deposit is determined by optical profilometry (TalySurf CCI). **Figure 2 (d)** reveals higher deposition of particles at the contact edge subjected to minimum confinement. The difference of deposition height between **Region I** and **Region II** for $A_{df}/A_v = 0.21$, $\Delta h$ is found $\sim 24\%$ for $L_{max}/L_{min} = 2$. Such deposit variations are minimized for low levels of confinement ($A_{df}/A_v = 0.04$), for the same level of asymmetry (Fig. 2c). It is obvious that even for high degree of asymmetry, the flow is effectively isotropic due to very low confinement (low accumulation of vapor).

In an effort to develop a simple theory on the evaporation lifetime of a 2-D droplet array, we first appeal to the case of a single droplet. Continuous diffusion of vapour from the liquid-air interface ($c_s$) to the ambient ($c_\infty$) governs the evaporation process in an isolated sessile droplet. However, for a confined droplet, the presence of surrounding obstructive entities leads to vapour confinement. Consequently, a proximal region (accumulation length, $\overline{L_a}$ ) with enhanced vapour concentration ($c'_\infty > c_\infty$) is formed **(Bansal et. al. 2017b)**. In a two-dimensional array (as in the present case), individual droplets create a vapour confinement to their vicinal droplets and vice-versa. In a quasi-steady state, the water vapour diffuses in the direction perpendicular to the array plane (**Fig.3(a)**). Once the geometric area of the confinement is characterized using the Voronoi tessellation, the evaporation model of the system can be modeled in a simple manner.

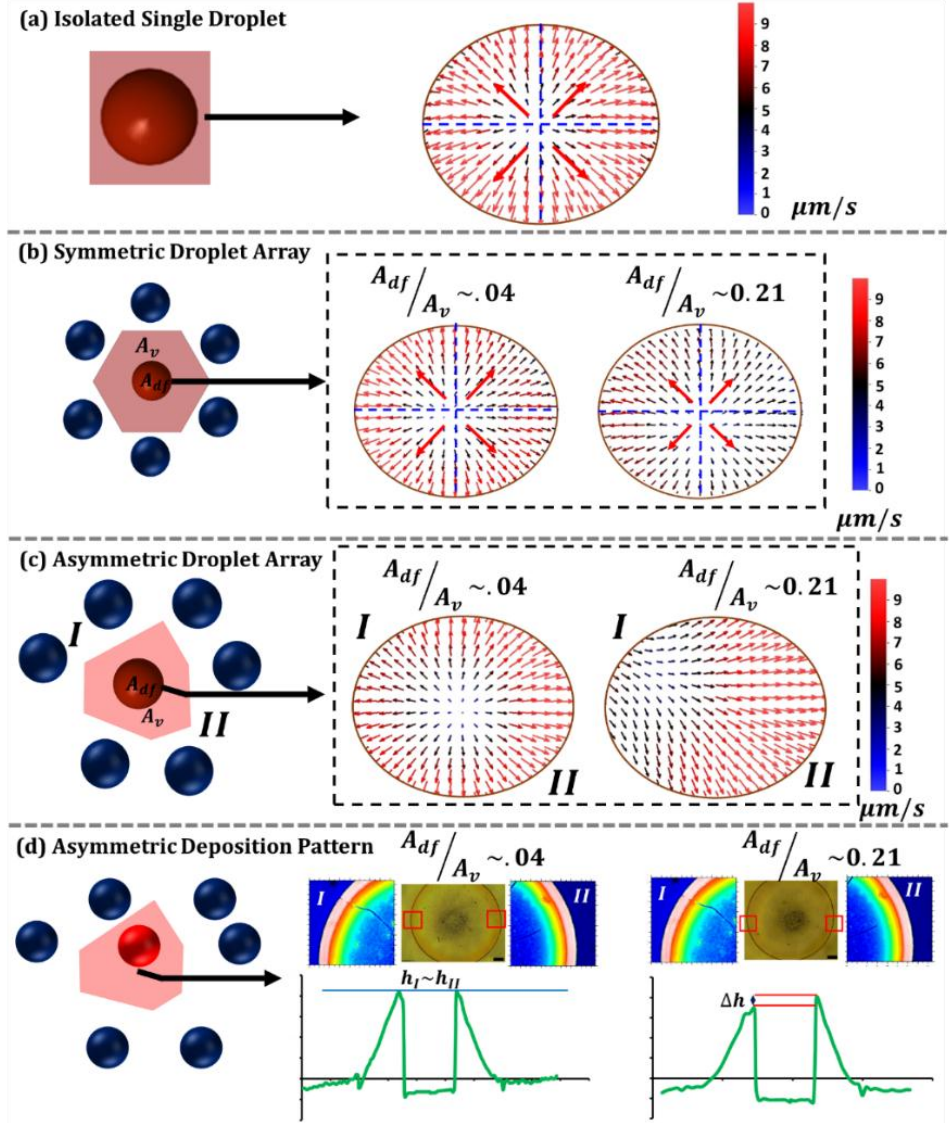

**Figure 2. Bottom view internal flow dynamics. (a)** of an isolated sessile droplet. **(b)** A symmetric arrangement, $L_{max}/L_{min} = 1$ **(c)** An asymmetric arrangement, $L_{max}/L_{min} = 2$ **(d)** Coffee deposition pattern for asymmetric condition, $L_{max}/L_{min} = 2$.

For an isolated unconfined single droplet, the instantaneous rate of volume decay is given by the Fick's law; $\frac{dV}{dt} = \frac{-J_{avg}A_d}{\rho} = \frac{-2\pi DMRf(\theta)(c_s-c_\infty)}{\rho}$, $V$ is the instantaneous droplet volume, M is molecular mass, $J_{avg}$ is the global evaporation flux value averaged around the droplet surface, $A_d$ is the instantaneous droplet surface area, ρ is the density of the working fluid, $D$ is the diffusion constant of vapor in air, $R$ and $\theta$ are the instantaneous droplet contact radius and contact angle respectively, and $c_s$ and $c_\infty$ are the vapor concentration near and far away from the liquid-air interface, respectively. Note that here we have neglected the convective effects arising due to concentration difference of water vapour at the droplet surface and the ambient (Grashof $Gr_s < 1$, for details see **Figure S3**, **Supplementary file**).

For $\theta_c > 10°$, $f(\theta) = (0.00008957 + 0.633\theta + 0.116\theta^2 - 0.08878\theta^3 + 0.01033\theta^4)/\sin\theta$ **(Picknett & Bexon, 1977; Erbil, 2012; Hu & Wu, 2016)**. The average rate of volume decay for an unconfined droplet is given as $\frac{dV}{dt}\big|_{uc,avg} = A\frac{dV}{dt}\big|_{t=0} = A\frac{-2\pi DMR_i f(\theta_i)(c_s-c_\infty)}{\rho}$, the subscripts $uc$ and $avg$ refer to unconfined and average, respectively. Here, $R_i$ and $\theta_i$ are the initial droplet contact radius and contact angle respectively, and A is a constant $(\sim O(1))$.

In the present case of 2-D arrangement, the interstitial/void region between the neighboring droplets accumulates the vapor diffused from the liquid-air interface of individual droplets. The evaporation process of a confined droplet, in effect, is a two step-process. The vapour diffusion from the droplet surface ($c_s$) to the confinement ($c'_\infty$) constitutes the first step and is followed by the vapour diffusion from the confinement to the ambient ($c_\infty$). Therefore, for the first step,

$$\frac{dV}{dt}\bigg|_{c,avg} = A\frac{-2\pi DMR_{ci}f(\theta_{ci})(c_s - c'_\infty)}{\rho} \quad (1)$$

subscript $c$ represents the confined droplet.

For the second step

$$\frac{dV}{dt}\bigg|_{avg} = \frac{-DMA_v(c'_\infty - c_\infty)}{\rho \overline{L_a}} \quad (2)$$

Here we introduce the accumulation length, $\overline{L_a}$, which signifies the length scale at which the vapor field relaxes to the ambient condition (**Supplementary Figure S4(a)**). At equilibrium, $\overline{L_a}$ can be estimated by equating (1) and (2),

$$\overline{L_a} = \frac{A_v(c'_\infty - c_\infty)}{2\pi AR_{ci}f(\theta_{ci})(c_s - c'_\infty)} \quad (3)$$

From (1) and (2),

$$(c_s - c'_\infty) = \frac{(c_s - c_\infty)}{1 + (2\pi AR_{ci}f(\theta_{ci})\overline{L_a}/A_v)} \quad (4)$$

Substituting (4) in (1), we get an expression relating the average volumetric decay of a confined and unconfined droplet as:

$$\frac{dV}{dt}\bigg|_{c,avg} = \frac{dV}{dt}\bigg|_{uc,avg}\left[\frac{1}{1 + (2\pi AR_{ci}f(\theta_{ci})\overline{L_a}/A_v)}\right] \quad (5)$$

Since $\frac{dV}{dt}\big|_{c,avg} \sim \frac{V_d}{t_c}$ and $\frac{dV}{dt}\big|_{uc,avg} \sim \frac{V_d}{t_{uc}}$,

$$\frac{t_c}{t_{uc}} = 1 + \left[\frac{2\pi AR_{ci}f(\theta_{ci})\overline{L_a}}{A_v}\right] \quad (6)$$

here $V_d$ is the initial droplet volume, $t_c$ and $t_{uc}$ are the total lifetimes of confined and unconfined droplets, respectively. In the current arrangement, as the top-view of the central droplet is easily accessible, we cast the lifetime scaling as the function of the droplet frontal area from the top-view ($A_{df}$). Here, we need to elucidate the droplet projection (as seen from the top) from the vantage of different substrates. For hydrophilic substrates, the projected droplet diameter and the contact diameter are the same, hence, $A_{df,hydrophilic} = \pi R_{ci}^2$ (**Supplementary Figure S4(b)**). On the contrary, for the hydrophobic substrates the projected diameter is equivalent to the droplet meridional diameter ($D$), given as $2R_{ci}cosec(\theta_{ci})$. Therefore, for the hydrophobic substrates; $A_{df,hydrophobic} = \pi R_{ci}^2 cosec(\theta_{ci})^2$ (**Supplementary Figure S4(b)**). Droplet confinement in the x-y plane restricts the available length scale for vapor diffusion in the z-direction (**Figure S4(b)**). Therefore, $\overline{L_a}$ can be scaled as $\beta R_{ci}$, and $\beta \sim 4$ (**Supplementary Figure S5**).

For hydrophilic substrates

$$\left.\frac{t_c}{t_{uc}}\right|_{hydrophilic} = 1 + \left[\frac{2A\beta\pi R_{ci}^2 f(\theta_{ci})}{A_v}\right] = 1 + \left[2A\beta f(\theta_{ci})\frac{A_{df,hydrophilic}}{A_v}\right] \quad (7)$$

For hydrophobic substrates

$$\left.\frac{t_c}{t_{uc}}\right|_{hydrophobic} = 1 + \left[\frac{2A\beta\pi R_{ci}^2 cosec(\theta_{ci})^2 f(\theta_{ci})}{A_v cosec(\theta_{ci})^2}\right] \quad (8)$$

$$= 1 + \left[\frac{2A\beta f(\theta_{ci})}{cosec(\theta_{ci})^2}\frac{A_{df,hydrophobic}}{A_v}\right]$$

Equations (7) and (8) present a predictive formulation for the evaporation lifetime scaling of sessile droplets confined in a two-dimensional array. The formulation is a generalized two-dimensional complete representation as compared to existing specialized mathematical expressions **(Hatte et. al. 2019; Bansal et. al. 2017a; Bansal et. al. 2017b)**.

The theoretical formulation shows linear dependence of evaporation lifetime of confined droplets on a parameter based on the initial contact angle, i.e. $f(\theta_{ci})$ and $cosec(\theta_{ci})$, which is known *a priori,* and the relative proximity of the neighboring droplets. Therefore, the scaling argument is uniquely applicable to a variety of droplet-substrate combinations in addition to symmetric/asymmetric arrangements. The universality of the theoretical formulation is also tested across different droplet-substrate combinations (GDL, glass, PDMS: **Fig. 3c**). The experimental data shows excellent agreement with the theoretical formulation (**Figure 3(c)**) across a wide range of governing parameters especially low to high confinements.

The experimental data (raw data shows variation of two times; **Figure S2**) collapse into a universal trend suggests that equations 7 and 8 can be used uniquely to predict the evaporation lifetime of any droplet in a 2D array for any droplet-substrate combinations. Note that we have also shown the data spread with respect to highest (GDL substrate) and lowest (Glass substrate) bounds of the theoretical trends. Furthermore, experiments have also been carried out on some readily available substrates such as; Packaging plastic ($\theta \sim 65° \pm 2°$), Plastic Book Cover ($\theta \sim 35° \pm 2°$), Borosil ($\theta \sim 64° \pm 2°$), Silicone-Wafer($\theta \sim 72° \pm 2°$), and Duct tape ($\theta \sim 44° \pm 2°$) **(Supplementary Figure S6)**. The experimental results reside well in the limiting theoretical lines i.e. between the hydrophilic and hydrophobic ranges (**Figure 3(a)**). The inclusiveness of the presented theoretical analysis for *particle laden droplets* has also been experimentally validated. Droplets of aqueous suspension of silica (SiO$_2$) nanoparticles at the particle loading by weight varying between w=0.5% (TM0.5, dilute loading) to w=5% (TM5, dense loading) are considered. The scaling of the evaporation lifetime for nano-suspension droplets is found to be accommodative irrespective of the loading rate **(Figure 3(a))**. This can be potentially exploited towards achieving a desired patterning of particles around the droplet by exclusive utilization of a symmetry breaking phenomenon in the flow profile due to cooperative evaporation. These outcomes signify the universality of the formulation which consolidates various combinations of liquid and substrate (with variation in their individual evaporation characteristics) into a global trend. It proves that irrespective of the surface topologies, hydrophobicity, and the droplet array configuration, the global trend of the scaling remains unaltered.

We have also compared the present findings with the analytical solutions of Carrier et al. **(2016)**. Their crude approximation of representing the vicinal droplets to super-drop fails for a sufficiently sparse (non-interacting) droplet arrangement and the scaling value overshoots to 2 (**Figure 3(b)**). Current analytical model when brought down to their specific case (drop arrangements on a polystyrene surface, Petri dishes, $\boldsymbol{\theta \sim 94°}$), it agrees well with the experimental results. More importantly, their overestimation is rectified in the present model. Detailed comparison with Carrier et al **(2016)** model can be found in **(Supplementary materials).** This outcome provides ample proof towards the predictive capability of the present formulation for any substrate with variable topography, chemical composition and interfacial interactions.

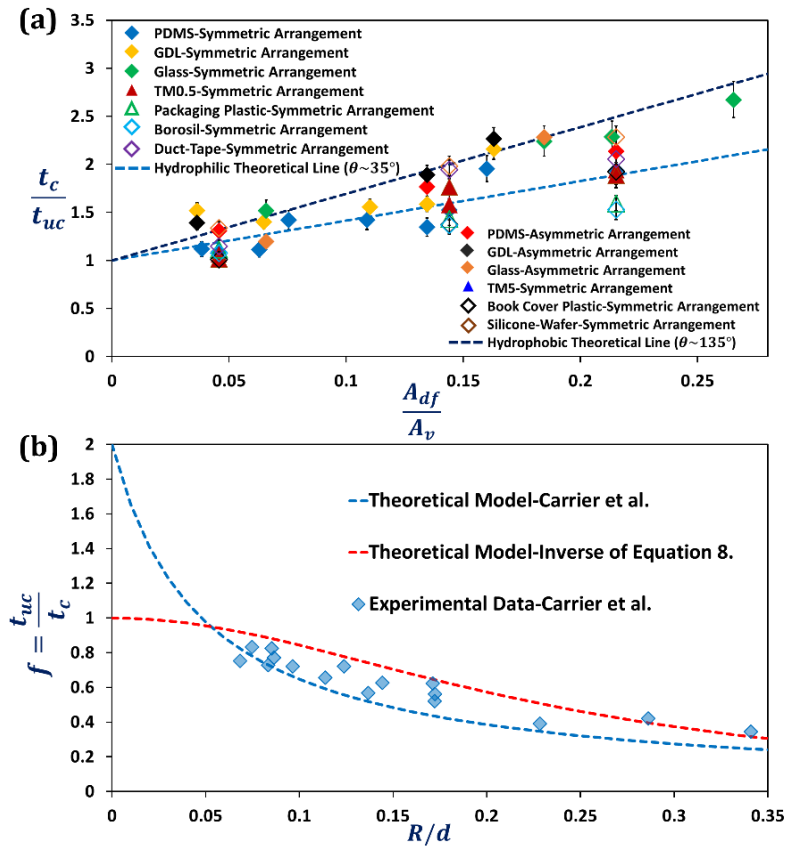

**Figure 3. (a) Evaporation lifetime scaling $\left(\frac{t_c}{t_{uc}}\right)$ as a function of $\left(\frac{A_{df}}{A_v}\right)$ (b) Comparison of the current model with the theoretical model of Carrier et al. (2016).**

To summarize, we have captured universalities in the cooperative and collective evaporation characteristics of an arbitrary two-dimensional droplet array. Our results not only provide a better understanding of the physics of droplet array evaporation, but also provide an extremely simple method to assemble diverse nonvolatile solutes into complex ordered structures. The process of cooperative evaporation is characterized by the geometric confinement created by the presence of multiple droplets in the vicinity. The method of Voronoi tessellation used in the present study extends the predictive capability of the analytical model to the case of random arrangement of sessile droplets. The incorporation of substrate wettability effects in the present model extends the predictive criteria of evaporation lifetime scaling to all possible combinations of working fluid and substrates, including the considerations of nanoparticle-laden droplets. Such a generalization in understanding the local and global aspects of evaporation, encompassing such diverse scenarios, may open up new vistas in industrial, biological and medical applications that have hitherto remained unaddressed.

*SB is grateful to the Swarnajayanti Fellowship, Department of Science and Technology, Govt. of India, for the financial support. SC acknowledges the financial support provided by the Department of Science and Technology, Govt. of India, by means of the J. C. Bose National Fellowship, for executing this research.*